\documentclass{article}%
\usepackage{amsfonts}
\usepackage{amsmath}
\usepackage{amssymb}
\usepackage{graphicx}%
\providecommand{\U}[1]{\protect\rule{.1in}{.1in}}
\begin{document}

\title{The effect of radiation corrections to the mass of an electron and positron on
the polarization operator of a photon in a magnetic field}
\author{V.M. Katkov\\Budker Institute of Nuclear Physics,\\Novosibirsk, 630090, Russia\\e-mail: katkov@inp.nsk.su}
\date{\ }
\maketitle

\begin{abstract}
The polarization operator of a photon in a constant and uniform magnetic field
is studied taking into account the radiation width and shift of the Landau
levels in both weak and strong fields compared with the critical field
$H_{0}=4,41\cdot10^{13}$ $%
%TCIMACRO{\unit{G}}%
%BeginExpansion
\operatorname{G}%
%EndExpansion
$. A general formula is obtained for the polarization operator of a photon in
which radiation effects are taken into account. Now diverging previously
threshold terms have a finite value. The conditions are formulated under which
the energy levels completely overlap, and thereby the most appropriate
application of the semiclassical operator method to the problem under study becomes.

\end{abstract}

\subsection{Introduction}

The study of QED processes in high magnetic field close to and above the
critical field strength $H_{0}=m^{2}/e=4,41\cdot10^{13}$ $%
%TCIMACRO{\unit{G}}%
%BeginExpansion
\operatorname{G}%
%EndExpansion
$ (the system of units $\hbar=c=1$ is used ) is of interest primary because
such fields exist in nature. It is commonly accepted that the magnetic field
of neutron stars (pulsars) reaches $H\sim10^{11}\div10^{13}%
%TCIMACRO{\unit{G}}%
%BeginExpansion
\operatorname{G}%
%EndExpansion
$ \cite{[1]}. This range of magnetic fields is obtained in the rotating
magnetic dipole model, where a pulsar loses energy through magnetic dipole
radiation. Predictions of this model are in good agreement with the
observation of radio radiation of pulsars. Several thousand radio pulsars are
currently known. Another class of neutron stars called magnetars \cite{[2]},
was discovered when observing X and gamma rays. In this case, the existing
models give much higher magnetic fields $H\sim10^{14}\div10^{15}$ $%
%TCIMACRO{\unit{G}}%
%BeginExpansion
\operatorname{G}%
%EndExpansion
$. In view of this circumstance, it is interest to describe the motion of a
photon and electron (positron) in fields both above and below the critical
field. This motion is accompanied by the photon conversion into a pair of
charged particles when the transverse photon momentum is larger than the
process threshold value $k_{\perp}>2m.$ When the field change is small on the
characteristic length of process formation (for example, when this length is
smaller then the scale of heterogeneity of the neutron star magnetic field),
the consideration can be realized in the constant field approximation. The
polarization operator of the photon in a constant and uniform electromagnetic
field of any configuration was obtained for the first time in 1971 by Batalin
and Shabad \cite{[3]} who used the Green function found by Schwinger
\cite{[4]}. Such calculation fore a pure magneic field were carried out in
1974 by Tsai \cite{[5]}. The singular behavior of the polarization operator
near the electron-positron pair production thresholds was analyzed in 1975 by
Shabad \cite{[6]}. In 1975 the contribution of charged-particles loop in an
electromagnetic field with $n$ external photon lines had been calculated in
\cite{[7]}. For $n=2$ the explicit expressions for the contribution of scalar
and spinor particles to the polarization operator of photon were given in this
work. For the contribution of spinor particles obtained expressions coincide
with the result of \cite{[3]}, but another form is used.

The polarization operator in a constant magnetic field has been investigated
well enough in the energy region lower and near the pair creation threshold
(see, for example, the papers \cite{[8],[9]} and the bibliography cited there.
In \cite{[10]} a general spectral-integral formula was obtained in which
diverging threshold terms were identified in an explicit analytical form. The
remaining members are presented in a form very convenient for numerical calculations.

In this paper, we consider radiation corrections to the energy of an electron
(positron) in a magnetic field and their effect on the polarization operator
of a photon. Such an effect is most noticeable at superstrong fields and
photon energies near the first excited thresholds for the production of
electron-positron pairs. The expression for the mass operator of an electron
(positron) in the $a$-order in a constant magnetic field was obtained in
\cite{[11], [12]}. It is a twofold integral of a surprisingly compact
expression containing, however, oscillating functions. In \cite{[13]}, a
detailed study of the mass operator and the radiation intensity (obtained in
this work) was carried out for arbitrary values of the magnetic field $H$ and
electron energy $\varepsilon$. We will be especially interested in the results
obtained for superstrong fields $H\gg H_{0}$ and energies at the first Landau
levels $n=0$ and $n\gtrsim1$.

\section{ Radiation Width and Shift of the Landau Levels}

The general expression for the mass operator of an electron in a constant and
uniform magnetic field, having a diagonal form, can be found in \cite{[13]}:%

\begin{align}
M  &  =\frac{\alpha m}{2\pi}\int_{0}^{\infty}\frac{dx}{x}\int_{0}%
^{1}due^{-\mathrm{i}ux/2\mu}\{\frac{1}{\Delta}\mathrm{\exp}[2\mathrm{i}%
n(a(x)-\frac{ux}{2})]\label{1}\\
&  [(\rho(1-u)-u)((c(x)+u(s(x)-c(x)))-us(x)+\\
&  \mathrm{i}\zeta\gamma_{\perp}u(1-u)x(\frac{c(x)}{x^{2}}-1+s(x))+1+u]-1-u\}.
\end{align}
Here%

\begin{align}
\rho &  =2n\mu=\gamma^{2}-1,\ \ \ \ \mu=H/H_{0},\ \ \ n=n_{0}+(1-\zeta
)/2,\label{2}\\
a(x) &  =\arctan\frac{uc(x)}{x\left(  1-us(x)\right)  },\ \ \ c(x)=1-\cos
(x),\ s(x)=1-\frac{\sin(x)}{x},\\
\zeta &  =\pm1(n\geq1),\ \ \ \ \ \ \zeta=1(n=0),\\
\Delta &  =1-2u(1-u)s(x)+u^{2}(2c(x)/x^{2}-1).
\end{align}
In the limit of the superstrong field ($\mu\gg1$), in \cite{[13]}, the
following expression was obtained with power accuracy for the correction to
the electron mass at $n=0$:%

\begin{equation}
M_{0}=\frac{\alpha m}{4\pi}\left[  (\ln2\mu-C-3/2)^{2}+A+O(1/\mu)\right]
,\ \ \ A=4,02816.... \label{31}%
\end{equation}
Here $C=0,577...$is Euler constant. For large $\mu$, the following approximate
expression for the radiation corrections to the electron energy at $n\geq1$
was obtained in \cite{[13]}:%
\begin{align*}
\varepsilon_{n}^{r}  &  =\frac{\alpha eH}{2\varepsilon_{n}}\left(
-\mathrm{i}\delta_{1n}+\delta_{2n}\right)  ,\ \ \ \varepsilon_{n}\simeq
\sqrt{2eHn};\\
\delta_{1n}  &  \simeq a_{1}n^{1/3}-k_{0}+a_{2}n^{-1/3}+\zeta(a_{3}%
n^{1/6}+a_{4}n^{-1/2})/\sqrt{\mu},\\
a_{1}  &  =1,8401;\ k_{0}=1,5213;\ a_{2}=0,3180;\ a_{3}=0,1606;\ a_{4}%
=0,2593;\\
\delta_{2n}  &  =b_{1}n^{1/3}-\kappa_{0}+b_{2}n^{-1/3}+\\
&  \zeta\lbrack-b_{3}n^{1/6}+\frac{1}{\pi\sqrt{2n}}(\ln\mu+c_{1}\ln
n-c_{2})]/\sqrt{\mu};
\end{align*}%
\begin{equation}
\ \ \ \ b_{i}=a_{i}/\sqrt{3},\ \ \ \kappa_{0}=0,5690/2\pi,\ \ c_{1}%
=11/30,\ \ c_{2}=0,3964. \label{4}%
\end{equation}
With good numerical accuracy, the formula (\ref{4}) can also be used at the
lowest excited levels at $n\sim1$ (see \cite{[13]}). For $\mu\gg1$, the
contribution of the terms $\varpropto\zeta$ is small and can be neglected. In
this case, the quantity $\varepsilon_{n}^{r}$ does not depend on the electron
mass. Leaving the main term of the expansion at $n\gg1$, we have%

\begin{equation}
\varepsilon_{r}=c\alpha\omega_{0}n^{1/3}(1/\sqrt{3}-\mathrm{i}),\ \ \ c=\frac
{7}{9}\Gamma\left(  \frac{2}{3}\right)  \left(  \frac{2}{3}\right)
^{1/3}\simeq0,920...,\label{5}%
\end{equation}
where $\omega_{0n}=eH/\varepsilon_{n}$. The formula (\ref{5}) is also
applicable for $H\ll H_{0}$ if the condition $eHn^{1/3}\gg m^{2}$ is
fulfilled. For $n\gg1$, the distance between the Landau levels is
$\delta\varepsilon_{n}\simeq\omega_{0n}$, and we have $\varepsilon_{n}%
^{r}/\delta\varepsilon_{n}\sim\alpha n^{1/3}.$ In weak fields, the formula
(\ref{5}) is applicable for $n$ $\gg(H_{0}/H)^{3}$ and then for $H/H_{0}%
\lesssim\alpha$ the energy levels completely overlap. It should be noted that
the condition $eHn^{1/3}\geq m^{2}$ is in fact a threshold for the creation of
pairs at these levels in weak fields. In this case, the energy of the emitted
photon is of the order of the entire electron energy $\varepsilon$. In the
opposite case $eHn^{1/3}\ll m^{2},$ the emitted harmonic is $\nu\sim\gamma
^{3}\ll n$, and relatively soft photons with a frequency $\omega\sim\nu
\omega_{0}\ll\varepsilon$ are emitted. In this case we have $\varepsilon
_{r}/\delta\varepsilon\sim\alpha\gamma$ and the energy levels completely
overlap for $\alpha\gamma\gtrsim1.$

\section{ Photon polarization operator}

In this section, we will use the results of \cite{[10]}. In a purely magnetic
field, we have in covariant form%

\begin{align}
\ \ \Pi^{\mu\nu}  &  =-\sum_{i=2,3}\kappa_{i}\beta_{i}^{\mu}\beta_{i}^{\nu
},\ \ \ \beta_{i}\beta_{j}=-\ \delta_{ij},\ \ \ \beta_{i}k=0;\label{6}\\
\beta_{2}^{\mu}  &  =(F^{\ast}k)^{\mu}/\sqrt{-(F^{\ast}k)^{2}},\ \ \ \beta
_{3}^{\mu}=(Fk)^{\mu}/\sqrt{-(Fk)^{2}},\label{7}\\
FF^{\ast}  &  =0,\ \ \ F^{2}=F^{\mu\nu}F_{\mu\nu}=2(H^{2}-E^{2})>0,
\end{align}
where $F^{\mu\nu}$ is the electromagnetic field tensor, $F^{\ast\mu\nu}$ is
the dual tensor, $k^{\mu}$ is the photon momentum, $(Fk)^{\mu}=$ $F^{\mu\nu
}k_{\nu}$. The real part $\kappa_{i}$ determines the refractive index of the
photon $n_{i}$ with polarization $e_{i}=\beta_{i}$:%

\begin{equation}
\ n_{i}=1-\frac{\mathrm{\operatorname{Re}}\kappa_{i}}{2\omega^{2}}. \label{9}%
\end{equation}
We will consider the process in the laboratory system, where the electric
field is absent, and the photon moves perpendicular to the magnetic field. In
this case%

\begin{equation}
r=\omega^{2}/4m^{2}. \label{10}%
\end{equation}
For $r>1$, the eigenvalues of the polarization operator contain the imaginary
part, which determines the probability of the production of an
electron-positron pair by a photon with polarization $\beta_{i}$:%

\begin{equation}
W_{i}=-\frac{1}{\omega}\mathrm{\operatorname{Im}}\kappa_{i}. \label{11}%
\end{equation}
The eigenvalues of the polarization operator $\kappa_{i}$ in the first order
of perturbation theory with respect to $\alpha$ contain root divergences for
$r=r_{lk}$, where%

\begin{align}
r_{lk}  &  =(\varepsilon_{l}+\varepsilon_{k})^{2}/4m^{2},\ \ \label{12}\\
\varepsilon_{l}  &  =\sqrt{m^{2}+2eHl}=m\sqrt{1+2\mu l}.
\end{align}

Consider the photon energy region near the ground state $r_{00}$. In the case
$\mu\gg1,\ \ |r-r_{00}|\ll1,$%

\begin{equation}
\kappa_{2}\simeq-\frac{4}{3\pi}\alpha m^{2},\ \ \ \ \kappa_{3}\simeq
\frac{\alpha}{\pi}eH[4-\frac{\mathrm{i}\pi}{\sqrt{(r-r_{00})}}]. \label{13}%
\end{equation}
From the formulas (\ref{31}) and (\ref{12}) it follows that%

\begin{equation}
r_{00}=(1+\delta_{0})^{2},\ \delta_{0}=M_{0}/m\ \ll1.\label{14}%
\end{equation}
It can be seen that the root divergence for the lower threshold is slightly
shifted to the point $r\simeq1+2\delta_{0}$. Another situation arises when
particles are born at higher levels. For example, for $r_{20}>$ $r>$ $r_{10}$
we have for the terms containing the root divergence
\begin{align}
\kappa_{2}^{10} &  =\alpha m^{2}\mu r\frac{2}{\pi}\exp\left(  -\frac{2r}{\mu
}\right)  \nonumber\\
&  \ \times\left[  \frac{\mu/2r-1}{\sqrt{h(r)}}A(r)-\frac{1}{2r}\ln
(\mu+1-r)\right]  ,\label{15}\\
\kappa_{3}^{10} &  =\alpha m^{2}\mu r\frac{2}{\pi}\exp\left(  -\frac{2r}{\mu
}\right)  \label{16}\\
&  \times\left[  \frac{\mu/2r-1-2/\mu}{\sqrt{h(r)}}A(r)-\frac{1}{2r}\ln
(\mu+1-r)+\frac{2}{\mu}\right]  ,\nonumber\\
A(r) &  =\arctan\frac{r-\mu/2}{\sqrt{h(r)}}+\arctan\frac{r+\mu/2}{\sqrt{h(r)}%
}\label{161}\\
&  =\pi-\arctan\frac{\sqrt{h(r)}}{r-\mu/2}-\arctan\frac{\sqrt{h(r)}}{r+\mu
/2},\nonumber\\
h(r) &  =(1+\mu)r-r^{2}-\mu^{2}/4.\label{162}%
\end{align}
For $r-$ $r_{10}<<1,$ $-h(r)\simeq\sqrt{1+2\mu}(r-r_{10})\ll1$, and the
expressions (\ref{15})-(\ref{16}) - have root divergence at $r=r_{10}$%

\begin{align}
\kappa_{i}^{10} &  \simeq-4\mathrm{i}\alpha m^{2}r\exp\left(  -\frac{2r}{\mu
}\right)  \frac{\beta_{i}}{\sqrt{-h(r)}},\label{17}\\
\beta_{2} &  =\frac{\mu}{2}-\frac{\mu^{2}}{4r},\ \ \beta_{3}=1+\frac{\mu}%
{2}-\frac{\mu^{2}}{4r}.\nonumber
\end{align}
We take into account that%

\begin{equation}
r-r_{10}\simeq2(\sqrt{r}-\sqrt{r_{10}})\sqrt{r_{10}}=\frac{1}{m}%
(\omega-\varepsilon_{0}-\varepsilon_{1})\sqrt{r_{10}},\label{20}%
\end{equation}
as well as the radiation width and shift of the first two Landau levels
(\ref{31})-(\ref{4}).Then for $\omega=\varepsilon_{0}+\varepsilon
_{1}+\mathrm{\operatorname{Re}}(\varepsilon_{0}^{r}+\varepsilon_{1}^{r})$ we
obtain the following expression for $\kappa_{i}^{10}$%

\begin{equation}
\kappa_{i}^{10}\simeq-\sqrt{\frac{\alpha}{\delta_{1}}}(2\mu)^{3/4}%
\mathrm{e}^{-2}(1+\mathrm{i})m^{2} \label{22}%
\end{equation}
The spectral part of the polarization operator, containing the root
divergences at the photon energy when the electron and positron are generated
at the Landau levels, is given in \cite{[10]}. Along with the numbers of
energy levels $l$ and $k$, we use the numbers $m=l+k,$ $n=l-k$ $(n\leq m)$. We have:%

\begin{align}
\kappa_{i}^{s}  &  =\alpha m^{2}\frac{r}{\pi}%
%TCIMACRO{\dsum \limits_{n=0}^{n_{\max}}}%
%BeginExpansion
{\displaystyle\sum\limits_{n=0}^{n_{\max}}}
%EndExpansion
\left(  1-\frac{\delta_{n0}}{2}\right)  T_{i}^{(ns)}=-\mathrm{i}\alpha
m^{2}\mu\mathrm{e}^{-\zeta}%
%TCIMACRO{\dsum \limits_{m\geq n}^{m=n_{\max}}}%
%BeginExpansion
{\displaystyle\sum\limits_{m\geq n}^{m=n_{\max}}}
%EndExpansion
(2-\delta_{n0})\frac{\zeta^{n}k!}{\sqrt{g}l!}\nonumber\\
&  \times\left[  1-\frac{1}{\pi}\left(  \arctan\frac{2\sqrt{-g}}{2r-\mu
n}+\arctan\frac{2\sqrt{-g}}{2r+\mu n}\right)  \right]  D_{i};\ \ \ \label{27}%
\\
D_{2}  &  =\left(  \frac{m\mu}{2}-\frac{n^{2}\mu^{2}}{4r}\right)  F+2\mu
l\vartheta(k-1)\left[  2L_{k-1}^{n+1}(\zeta)L_{k}^{n-1}(\zeta)-L_{k}^{n}%
(\zeta)L_{k-1}^{n}(\zeta)\right]  ,\nonumber\\
D_{3}  &  =\left(  1+\frac{m\mu}{2}-\frac{n^{2}\mu^{2}}{4r}\right)  F+2\mu
l\vartheta(k-1)L_{k}^{n}(\zeta)L_{k-1}^{n}(\zeta),\nonumber\\
F  &  =\left[  L_{k}^{n}(\zeta)\right]  ^{2}+\vartheta(k-1)\frac{l}{k}\left[
L_{k-1}^{n}(\zeta)\right]  ^{2},\ \ \ \zeta=\frac{2r}{\mu}, \label{28}%
\end{align}
where $L_{k}^{n}(\zeta)$ is the generalized Laguerre polynomial.%

\begin{align}
g_{lk}(r)  &  =r^{2}-(1+m\mu)r+n^{2}\mu^{2}/4,\label{30}\\
n_{\max}  &  =[d(r)],\ \ \ \ \ d(r)=2(r-\sqrt{r})/\mu.
\end{align}
Near $r_{lk}$ $(r-r_{lk}\ll1)$, $g_{lk}(r)$ has the form
\begin{align}
g_{lk}(r)  &  \simeq(r-r_{lk})\sqrt{1+2(l+k)\mu+4kl\mu^{2}},\label{32}\\
r-r_{lk}  &  \simeq\left(  \frac{\omega}{2m}-\frac{\varepsilon_{l}%
+\varepsilon_{k}}{2m}\right)  \frac{\varepsilon_{l}+\varepsilon_{k}}{m}.
\label{33}%
\end{align}

We take into account the radiation corrections to the energy levels
(\ref{31})-(\ref{4}) and choose (taking into account the level shift) the
photon energy equal to the sum of the energies of the electron and positron.
Then%
\begin{align}
\left(  \frac{\omega}{2m}-\frac{\varepsilon_{l}+\varepsilon_{k}}{2m}\right)
&  \rightarrow-\mathrm{i\operatorname{Im}}\frac{\varepsilon_{l}^{r}%
+\varepsilon_{k}^{r}}{2m}=\mathrm{i}\frac{\sqrt{\mu}}{4\sqrt{2}}\left(
\frac{\delta_{1l}}{\sqrt{l}}+\frac{\delta_{1k}}{\sqrt{k}}\right)
,\label{34}\\
g_{lk}^{r} &  \simeq\mathrm{i}\frac{\sqrt{\mu}}{4\sqrt{2}}\left(  \frac
{\delta_{1l}}{\sqrt{l}}+\frac{\delta_{1k}}{\sqrt{k}}\right)  (\sqrt{1+2\mu
l}+\\
&  \sqrt{1+2\mu k})\sqrt{1+2(l+k)\mu+4kl\mu^{2}}\label{35}%
\end{align}
For $\mu\gg1$, we will consider the function $g$ when one of the particles is
born at the ground level $k=0$. Then $m=n=l$%
\begin{equation}
g\simeq\frac{\mathrm{i}\alpha}{2\sqrt{2}}\mu^{3/2}l\delta_{1l},\ \ \ g^{-1/2}%
\simeq\frac{2^{1/4}\mu^{-3/4}}{\sqrt{\alpha l\delta_{1l}}}(1-\mathrm{i}%
)\label{36}%
\end{equation}
When $r-r_{10}<<1$ the main term in the sum is determined by the values $m=n$
$=l=1,$ $k=0,$ $D_{2}=$ $\beta_{2},$ $D_{3}=$ $\beta_{3},$ $g=-h,$ which is
consistent with the expression (\ref{22}).

In order to take into account the radiation corrections to the Landau levels
in the general formula (\ref{27}), we introduce a modified function%
\begin{equation}
g_{lk}^{r}(r)=\left(  1-\frac{\varepsilon_{l}^{r}+\varepsilon_{k}^{r}%
}{2m(\sqrt{r}-\sqrt{r_{lk}})}\right)  g_{lk}(r), \label{37}%
\end{equation}
and instead of function $g$ in (\ref{27}) we shall be used this function
$g^{r}$ (\ref{37}). The inclusion of radiation corrections in nonsingular
terms leads to small corrections $\sim\alpha$ to the polarization operator and
we neglect these corrections.

\section{Conclusion}

We examined the effect of radiation corrections to the mass of an electron
(positron) on the polarization operator of a photon in a constant and uniform
magnetic field. In relatively weak fields $H\ll H_{0}$, the actual pair
production threshold determines the criterion $Hn^{1/3}/H_{0}\gtrsim1.$ Since
the relative width of the energy levels is $\alpha n^{1/3}$, then for
$H/H_{0}\lesssim\alpha$ these levels overlap and the quasiclassical
consideration of the process is most adequate \cite{[14]}. Thus, a detailed
consideration of these issues is necessary in the field of fields close to or
exceeding the critical Schwinger field. It is the existence of such fields in
pulsars and magnetars that was discussed in the Introduction. We have
formulated the procedure for incorporating radiation corrections into
previously obtained formulas for the polarization operator \cite{[10]}. As a
result, the expressions obtained do not contain root divergences in the
probability of pair production at the Landau levels.

\end{document}